\documentclass[fleqn,twoside]{article}
\usepackage[headings]{espcrc2}

\readRCS
$Id: espcrc2.tex,v 1.2 2004/02/24 11:22:11 spepping Exp $
\usepackage{graphicx, balance}
\usepackage[cmex10]{amsmath}
\usepackage{algorithmic}
\usepackage[dvipsnames,usenames]{color}
\usepackage{array}     	
\usepackage{subfigure}
\usepackage[normalem]{ulem}  
\usepackage[ruled, vlined]{algorithm2e}
\usepackage{verbatim}  	
\usepackage{tabularx}  	
\usepackage{multirow}  	
\usepackage{booktabs} 	
\usepackage[table]{xcolor}  
\usepackage[bookmarks]{hyperref}
\usepackage{url}
\usepackage[figuresright]{rotating}
\usepackage{ctable} 

\title{\textbf{Two-Hop Routing with Traffic-Differentiation for QoS Guarantee in Wireless Sensor Networks}}

\author{T Shiva Prakash\address[DCSE]{Department of Computer Science and Engineering, University Visvesvaraya College of\\~ Engineering, Bangalore University, Bangalore 560 001 India, Contact: spt@ieee.org\\},
K B Raja\addressmark,
K R Venugopal\addressmark,
S S Iyengar\address[DCSE]{Florida International University, Miami, Florida, USA\\},
L M Patnaik\address{Honorary Professor, Indian Institute of Science, Bangalore 560 012, India}}

\runtitle{Two-Hop Routing with Traffic-Differentiation for QoS Guarantee in Wireless Sensor Networks}
\runauthor{Shiva Prakash T, et al.,}

\begin{document}
\begin{abstract}

This paper proposes a Traffic-Differentiated Two-Hop Routing protocol for Quality of Service (QoS) 
in Wireless Sensor Networks (WSNs).
It targets WSN applications having different types of data traffic with several priorities.
The protocol achieves to increase Packet Reception Ratio (PRR) and reduce end-to-end delay while considering 
multi-queue priority policy, two-hop neighborhood information, link reliability and power efficiency. The protocol is modular and 
utilizes effective methods for estimating the link metrics. Numerical results show 
that the proposed protocol is a feasible solution to addresses QoS service differentiation for traffic with different priorities. 
\end{abstract}

\maketitle

\section{Introduction}
\label{section:Introduction}
Wireless Sensor Networks (WSNs) form a framework to accumulate and analyze real time data in smart environment applications. WSNs are composed of inexpensive low-powered micro sensing devices called $motes$[1], having limited computational capability, memory size, radio transmission range and energy supply.
These sensors are spread in an environment without any predetermined infrastructure and cooperate to accomplish common monitoring tasks which usually involves sensing environmental data. With WSNs, it is possible to assimilate a variety of physical and environmental information in near real time from inaccessible and hostile locations. 
\newcommand {\otoprule}{\midrule [\heavyrulewidth]}  
\begin{table*}
\centering
\caption{Our Results and Comparison with Previous Results for Differentiated QoS Routing in Wireless Sensor Networks.}
\begin{center}
\scriptsize	
    \begin{tabular}{lllllll}
    \hline
\bfseries {Related} & \bfseries {Protocol} & \bfseries {Considered} & \bfseries {Estimation} & \bfseries {Traffic} & \bfseries {Duplication} \\ 
\bfseries {Work}    &  \bfseries {Name}    & \bfseries {Metrics}    & \bfseries {Method} & \bfseries {Differentiation} & \bfseries \\
\hline
    E. Felemban & MMSPEED &one-hop delay, &EWMA &Delay &Towards \\
    \emph{et al.,}[2] &(Multi-path &link reliability & &requirement &the same \\
    &and Multi-SPEED &and residual & & &sink \\
    &Routing Protocol) &energy. & & & &\\
\hline
    M.M. & DARA (Data  &one-hop delay &Variance-based &Critical &Towards \\
    Or-Rashid  &Aggregate &and transmission & &Traffic and &different \\
    \emph{et al.,}[3] &Routing  &power. & &Non-Critical &sinks \\
    &Algorithm) & & &Traffic &\\
\hline
    Y. Li & THVR (Two-Hop &two-hop delay &WMEWMA &No &No \\
    \emph{et al.,}[4] &Velocity Based &and residual & & & \\
    &Routing Protocol) &energy. & & & \\
\hline
    D. Djenouri & LOCALMOR &one-hop delay &EWMA and &Regular, &Towards \\
    \emph{et al.,}[5] &(Localized &link reliability &WMEWMA &Reliability \\
    &Multi-objectives &residual energy & &-sensitive,Delay, &sinks \\
    &Routing) &and transmission & &-sensitive and & \\ 
    & &power. & &Critical Traffic& \\ 
\hline
    This paper &TDTHR &two-hop delay, &EWMA and &Regular, &Towards \\  
    &(Traffic &link reliability &WMEWMA &Reliability &different \\
    &-Differentiated &residual energy & &-Responsive,Delay, &sinks \\
    &Two-Hop  &and transmission & &-Responsive and & \\
    &Routing) & power. & &Critical Traffic & \\
\hline
    \end{tabular}
\end{center}
\end{table*}
\vskip 2mm
WSNs have a set of stringent QoS requirements that include timeliness, high reliability, availability and integrity. 
Various performance metrics that can be used to justify the quality of service include, packet reception ratio (PRR), defined as the probability of successful delivery should be maximized. 
The end-to-end delay which is influenced by the queuing delay at the intermediate
nodes and the number of hops traversed by the data flows of the session from the source to the receiver.
Sensor nodes typically use batteries for energy supply. Hence, energy efficiency and load balancing form important objectives while designing protocols for WSNs. 
Therefore, providing corresponding traffic differentiated QoS in such scenarios pose a great challenge. Our proposed protocol is motivated primarily by the deficiencies of the previous works (explained in the Section 2) and aims to provide better Quality of Service.
\vskip 2mm
This paper explores the idea of incorporating QoS parameters in making routing decisions the protocol proposes the following features.
\begin{enumerate}
\item
Data traffic is split into regular traffic with no specific QoS requirement, reliability-responsive traffic; which should be transmitted without loss but can tolerate some delay, delay-responsive traffic; which should be delivered within a deadline but may tolerate moderate packet loss and critical traffic; which has high significance and demanding both high reliability and short delay.
\item
Link reliability is considered while choosing the next router, this selects paths which have higher probability of successful delivery.
\item
Routing decision is based on two-hop neighborhood information and dynamic velocity that can be modified according to the required deadline, this results in significant reduction in end-to-end PRR.
\item
Choosing nodes with higher residual energy and minimum transmission power, balances the load among nodes 
and results in prolonged lifetime of the network.
\end{enumerate}

We test the performance of our proposed approaches by implementing our algorithms using $ns$-$2$ simulator. Our results
demonstrates the performance and benefits of TDTHR over earlier algorithms.
\vskip 2mm
The rest of the paper is organized as follows: Section 2 gives a review of related works. Section 3 and Section 4 explain the network model, notations, assumptions and working of the algorithm. Section 5 is devoted to the simulation and evaluation of the algorithm. Conclusions are presented in Section 6.

\section{Related Work}
\label{section:related work}
Stateless routing protocols which do not maintain per-route state is a favorable approach for WSNs. The idea of stateless routing is to use location information available to a node locally for routing, $i.e.$, the location of its own and that of its one-hop neighbors without the knowledge about the entire network. These protocols scale well in terms of routing overhead because the tracked routing information does not grow with the network size
or the number of active sinks. Parameters like distance to sink, energy efficiency and data aggregation, need to be considered to select the next router among the one-hop neighbors.
\vskip 2mm
SPEED (Stateless Protocol for End-to-End Delay)[6] is a well known stateless routing protocol for real-time communication in sensor networks. It is based on geometric routing protocols
such as greedy forwarding GPSR (Greedy Perimeter State Routing) [7][8]. 
MMSPEED (Multi-path and Multi-SPEED Routing Protocol) [2] is an extension of SPEED that focuses on differentiated QoS options for real-time applications with multiple different deadlines. It provides differentiated QoS options both in timeliness domain and the reliability domain. For timeliness, multiple QoS levels are supported by
providing multiple data delivery speed options. For reliability, multiple reliability requirements are supported by probabilistic multi-path forwarding. The protocol provides end-to-end QoS provisioning by employing localized geographic forwarding using immediate neighbor information without end-to-end path discovery and maintenance. It utilizes dynamic compensation which compensates for inaccuracy of local decision as a packet travels towards its destination. The protocol adapts to network dynamics. MMSPEED does not include energy metric during QoS route selection.  

\vskip 2mm
Sharif \emph{et al.}, [9] presented a new transport layer protocol that prioritizes sensed information based on its nature while simultaneously supporting the data reliability and congestion control features. Rusli \emph{et al.}, [10] propose an analytical framework model based on Markov Chain of OR and M/D/l/K queue to measure its performance in term of end-to-end delay and reliability in WSNs. 
\vskip 2mm
Koulali \emph{et al.}, [11] propose a hybrid QoS routing protocol for WSNs based on a customized Distributed Genetic Algorithm (DGA) that accounts for delay and energy constraints. Yunbo Wang \emph{et al.}, [12] investigate the end-to-end delay distribution, they develop a comprehensive cross-layer analysis framework, which employs a stochastic queuing model in realistic channel environments. Ehsan \emph{et al.}, [13] propose energy and cross-layer aware routing schemes for multichannel access WSNs that account for radio, MAC contention, and network constraints. Geographical routing protocols have been proposed, such as GREES (Geographic Routing with Environmental Energy Supply) [14], DHGR (Dynamic Hybrid Geographical Routing) [15], and EAGFS (Energy Aware Geographical Forwarding Scheme) [16].
\vskip 2mm
DARA (Distributed Aggregate Routing Algorithm) [3] considers reliability, delay, and residual energy 
in the routing metric, and defines two kinds of packets: critical and non-critical packets. The same weighted metric is used for both types of packets, where the only difference is that a set of candidates reached with a higher transmission power is considered to route critical packets. For delay estimation, the authors use queuing theory and suggest a method that, in practice, needs huge amount of sample
storages.
\vskip 2mm
All the above routing protocols are based on one-hop neighborhood information. However, it is expected that multi-hop information can lead to improved performance in many issues including message broadcasting and routing. 
Chen \emph{et al.}, [17] study the performance of 1-hop, 2-hop and 3-hop neighborhood information based
routing and propose that gain from 2-hop to 3-hop is relatively minimal, while that from 1-hop to 2-hop based routing is significant.
Li \emph{et al.},[4] have proposed a Two-Hop Velocity Based Routing Protocol (THVR). 
Shiva Prakash  \emph{et al.}, [18] propose a Link Reliability based Two-Hop Routing protocol which achieves to 
reduce packet deadline miss ratio while considering link reliability, two-hop delay and power efficiency.
\vskip 2mm
Djenouri \emph{et al.}, [5] propose a new localized quality of service (QoS) routing protocol (LOCALMOR) it is based on differentiating QoS requirements
according to the data type, which enables to provide several and customized QoS metrics for each traffic category. With each packet,
the protocol attempts to fulfill the required data-related QoS metrics while considering power efficiency. The protocol proposed in this paper is different from LOCALMOR it considers two-hop transmission delay and queuing delay for selecting the optimal path.

\section{Problem Definition}
\label{section:pd}
The topology of a wireless sensor network may be described by a graph $G=(N,L)$, where $N$ is the set of nodes and $L$ is the set of links. The objectives are to,
\begin{itemize}
\item Maximize the Packet Reception ratio (PRR).
\item Reduce the end-to-end packet delay.
\item Improve the energy efficiency (ECPP-Energy Consumed Per Packet) of the network.
\end{itemize}
\subsection{Network Model and Assumptions}
In our network model, we assume the following:
\begin{itemize}
  \item The wireless sensor nodes consists of $N$ sensor nodes and a sink, the sensors are distributed randomly in a field. The nodes are aware of their positions through internal global positioning system (GPS).
  \item The $N$ sensor nodes are powered by a non-renewable on board energy source. All 
  nodes are supposed to be aware of their residual energy and have the same transmission power range.
  \item The sensors share the same wireless medium each packet is transmitted as a local broadcast in the neighborhood. We assume any MAC protocol, which ensures that among the neighbors in the local broadcast range, only the intended receiver keeps the packet and the other neighbors discard the packet.
  \item Like all localization techniques, [6][2][19] each node needs to be aware of its 
  neighboring nodes current state (ID, position, link reliability, residual energy etc), this is done via HELLO messages. 
  \item In addition, each node sends a second set of HELLO messages to all its neighbors informing them about its one-hop neighbors. 
  Hence, each node is aware of its one-hop and two-hop neighbors and their current state.
  \item Nodes are assumed to be stationary or having low mobility. The network density is assumed to be high enough to prevent the void situation.
\end{itemize}

\newcolumntype{A}{>{$}c<{$}}
\newcolumntype{B}{>{$}l<{$}}
\newcolumntype{C}{m{4.5cm}}
\begin{table}[t]
 		\caption{Notations Used in Section 4}
\scriptsize																										 
 	\begin{tabular}{BC}
    \toprule[0.5pt]   
    \multicolumn{1}{b{1.2cm}}{\bfseries Symbol}
    & \multicolumn{1}{b{3cm}}{\bfseries Definition}\\ 
    \toprule[0.5pt]   

    \raggedright N
    & Set of Nodes in the WSN\\

    \raggedright D
    & Destination Node \\
    
    \raggedright S
    & Source Node \\
    
    \raggedright dist(x,y)
    & Distance between a node pair $x,y$ \\
    
    \raggedright N_{1}(x)
    & Set of one-hop Neighbors of node $x$ \\

    \raggedright N_{2}(x)
    & Set of two-hop Neighbors of node $x$ \\
    
    \raggedright F_{1}^{+p}(x)
    & Set of node $x$'s one-hop favorable \\
    
    \raggedright
    & forwarders providing positive progress \\

    \raggedright
    & towards the destination D \\

    \raggedright F_{2}^{+p}(x,y)
    & Set of node $x$'s two-hop \\

    \raggedright
    & favorable forwarders \\

    \raggedright dt_{xy}
    & Estimated hop delay between $x$ and $y$ \\

    \raggedright dq_{x}
    & Estimated queuing delay at node $x$\\

    \raggedright t_{req}
    & Time deadline to reach Destination D\\
    
    \raggedright V_{req}
    & Required end-to-end packet delivery \\
    
    \raggedright
    & Velocity for deadline $t_{req}$ \\

    \raggedright V_{xy}
    & Velocity offered by $y$ $\in$ $F_{1}^{+p}(x)$  \\

    \raggedright V_{xy \rightarrow z}
    & Velocity offered by $y$ $\in$ $F_{2}^{+p}(x,y)$  \\

    \raggedright S_{req}
    & Node pairs satisfying $V_{xy\rightarrow z} \geq V_{req}$   \\
    
    \raggedright T_{a}(dist(x,y)^\alpha)
    & Transmission power cost from node x to node y\\
    
    \raggedright E_{y}
    & Remaining energy of node $y$\\
    
    \raggedright prr_{xy}
    & Packet Reception Ratio of link relaying \\

    \raggedright
    & node $x$ to node $y$ \\
    
    \raggedright \beta
    & Tunable weighting coefficient for $prr$ estimation\\
    
    \raggedright \gamma
    & Tunable weighting coefficient for queuing and \\

    \raggedright
    & transmission delay estimation\\
    
    \toprule[0.5pt] 
  \label{table:notations}
  \end{tabular}
\end{table}

\section{Algorithm}
\label{section:algo}
TDTHR has the following components: a link reliability estimator, a queuing and transmission delay estimator, a node forwarding metric incorporated with the two-hop dynamic velocity assignment policy and a queuing controller.
The proposed protocol LRTHR implements the modules for estimating queuing and transmission 
delay and packet delivery ratios using efficient methods. The packet delay is estimated at the node itself and the packet delivery ratio is estimated by the neighboring nodes. These parameters are updated on reception of a HELLO packet, the HELLO messages are periodically broadcast to update the estimation parameters. The overhead caused by the 1-hop and 2-hop updating are reduced by piggybacking the information in ACK, hence improving the energy efficiency. The notations used in this paper are given in Table \ref{table:notations}. The protocol is based on the following parameters: (i) Link Reliability Estimation; (ii) Queuing and Transmission Delay Estimation; (iii) Node Forwarding Metric; and (iv) Queuing Controller
\subsection{Link Reliability Estimation}
The Packet Reception Ratio (PRR) of the link relaying node $x$ to $y$ is denoted by $prr_{xy}$. It denotes the probability of successful 
delivery over the link. Window Mean Exponential Weighted Moving Average (WMEWMA) based link quality estimation is used for the proposed protocol. The window mean exponential weighted moving average estimation applies filtering on PRR, thus providing a metric that resists transient fluctuations of PRR, yet is responsive to major link quality changes.
This parameter is updated by node $y$ at each window and inserted into the HELLO message packet for usage by node $x$ in the next window. 
Eqn. \ref{eqn:Eq1} shows the window mean exponential weighted moving average estimation of the link reliability, $r$ is the number of packets received, $m$ is the number of packets missed and $\beta \in [0,1]$ is the history control factor, which controls the effect of the previously estimated value on the new one, $\frac{r}{r+m}$ is the newly measured PRR value.

\begin{equation}
prr_{xy} = \beta \times prr_{xy} + (1-\beta) \times \frac{r}{r+m} 
\label{eqn:Eq1}
\end{equation}
The PRR estimator is updated at the receiver side for each $w$ (window size) received packets, the computation complexity of
this estimator is $O(1)$. The appropriate values for $\beta$ and $w$ for a stable window mean exponential weighted moving average are $w=30$ and $\beta=0.6$\cite{EELRE1}.

\subsection{Queuing and Transmission Delay Estimation}
The nodal delay indicates the time spent to send a packet from node $x$ to its neighbor $y$, it is comprised of the 
queuing delay (delay$_Q$), contention delay (delay$_C$) and the transmission delay (delay$_T$).
\begin{equation}
delay_{node} = delay_{Q} + delay_{C} + delay_{T}
\label{eqn:Eq2}
\end{equation}
The queuing delay constitutes the time the packet is assigned to a queue for transmission and the time it starts being transmitted. During this time, the packet waits while other packets in the transmission queue are transmitted.
Every node evaluates its queuing delay $dq_x$ for the various classes of queues used, $i.e.,$ Critical-Queue, Delay-Responsive-Queue, Reliability-Responsive-Queue and Regular-Queue, each packet class has a different estimation of $dq_x$ for the queuing 
delay, $i.e.,$ $dq_x<packet.class>$. Eqn. \ref{eqn:Eq2a} shows the EWMA (Exponential Weighted Moving Average) update for queuing delay estimation, $dq$ is the current precise queue waiting time of the respective packet and $\gamma \in [0,1]$ is the 
tunable weighting coefficient.
\begin{equation}
\begin{split}
dq_{x}<packet.class> = &\gamma \times \\
 		       & dq_{x}<packet.class> \\
                       & + (1-\gamma) \times dq
\label{eqn:Eq2a}
\end{split}
\end{equation}
The transmission delay represents the time that the first and last bits of the packet
are transmitted. If $t_{s}$ is the time the packet is ready for transmission and becomes head of transmission queue, $t_{ack}$ the 
time of the reception of acknowledgment, $BW$ the network bandwidth and size of the acknowledgment then, $t_{ack} - sizeof(ACK)/BW - t_{s}$ is the recently
estimated delay. Eqn. \ref{eqn:Eq3} shows the EWMA (Exponential Weighted 
Moving Average) update for transmission delay 
estimation, which has the advantage of being simple and less resource demanding. 
\begin{equation}
\begin{split}
dt_{xy} = &\gamma \times dt_{xy} + (1-\gamma) \times \\
          &(t_{ack} - sizeof(ACK)/BW - t_{s})
\label{eqn:Eq3}
\end{split}
\end{equation}
$dt_{xy}$ includes estimation of the time interval from the packet that becomes head of line of $x$'s transmission queue until 
its reception at node $y$. This takes into account all delays due
to contention, channel sensing, channel reservation (RTS/CTS) if any, depending on the medium access control (MAC) protocol,
propagation, time slots etc. The computation complexity of the estimators are $O(1)$. The delay information is further exchanged among two-hop neighbors.

\begin{algorithm}
\begin{footnotesize}
\KwIn{$x$, $D$, $F_1^{+p}(x)$, $F_2^{+p}(x)$, $lt$}
\KwOut{Node $y$ providing positive progress towards D}
\BlankLine

$V_{req}$ = $\frac{dist(x,D)}{lt}$ \;
\For{each $y$ $\in$ $F_2^{+p}(x)$}{
	$V_{xy \rightarrow z}$ = $\frac{dist(x,D) - dist(k,D)}{dq_x<packet.class> + dt_{xy} + dq_y<packet.class> + dt_{yz}}$ \;
}
$S_{req}$ = \{$F_2^{+p}(x)$ : $V_{xy\rightarrow z} \geq V_{req}$\} \;
\If{($\mid$ $S_{req}$ $\mid$) = 1}{
	Return $y$ $\in$ $S_{req}$\;
}
\Else{
	\If{packet.class == delay.responsive}{
		\For{each $y$ $\in$ $S_{req}$}{
			Return $y$ with $max$ $E_y$ $min$ $T_a(dist(x,y)^\alpha$);
		}
	}
	\Else{
		\If{packet.class == critical}{
			\For{each $y$ $\in$ $S_{req}$}{
				$S_c$ = Return $y$ with $max$ $prr_{xy}$;		
			}
			\If{($\mid$ $S_{c}$ $\mid$) = 1}{
				Return $y$ $\in$ $S_{c}$\;
			}
			\Else{
				\For{each $y$ $\in$ $S_{c}$}{
					Return $y$ with $max$ $E_y$ $min$ $T_a(dist(x,y)^\alpha$);
				}
			}
		}
	}	
	
}
\caption{Traffic-Differentiated Two-Hop Routing (TDTHR)}
\label{algo:tdthr}
\end{footnotesize}
\end{algorithm}

\subsection{Node Forwarding Metric}
In the wireless sensor network, described by a graph $G=(N,L)$. If node $x$ 
can transmit a message directly to node $y$, the ordered pair is an element of $L$. We define for each node $x$ the set $N_1(x)$, which contains the nodes
in the network $G$ that are one-hop $i.e.,$ direct neighbors of $x$.
\begin{equation}
N_1(x) = \{y:(x;y) \in E \hspace{0.1cm} and \hspace{0.1cm} y \neq x\}
\label{eqn:Eq4}
\end{equation}

Likewise, the two-hop neighbors of $x$ is the set $N_2(x)$ $i.e.,$
\begin{equation}
N_2(x) = \{z:(y;z) \in E \hspace{0.1cm} and \hspace{0.1cm} y \in N_1(x), \hspace{0.1cm} z \neq x\}
\label{eqn:Eq5}
\end{equation}
The euclidean distance between a pair of nodes $x$ and $y$ is defined by $dist(x,y)$. We define $F_{1}^{+p}(x)$ as the set of $x$'s one-hop favorable forwarders providing positive progress towards the destination $D$. It consists of nodes that are closer to the destination than $x$, $i.e.,$
\begin{equation}
\begin{split}
F_1^{+p}(x) = &\{y \in N_1(x): \hspace{0.1cm} dist(x,D) \hspace{0.1cm} - \\
              &\hspace{0.1cm} dist(y,D) > 0\}
\label{eqn:Eq6}
\end{split}
\end{equation}
$F_{2}^{+p}(x)$ is defined as the set of two-hop favorable forwarders $i.e.,$
\begin{equation}
\begin{split}
F_2^{+p}(x)& = \{y\in F_1^{+p}(x),z\in N_1(y):\hspace{0.1cm} \\
           & \quad dist(y,D)\hspace{0.1cm}-\hspace{0.1cm} dist(z,D)>0\}
\label{eqn:Eq7}
\end{split}
\end{equation}
We define two velocities; the required velocity $V_{req}$ and the velocity offered by the two-hop favorable forwarding pairs. In SPEED[6], the velocity provided by each of the forwarding nodes in ($F_1^{+p}(x)$) is.
\begin{equation}
V_{xy} = \frac{dist(x,D) - dist(y,D)}{dt_{xy}}
\label{eqn:Eq8}
\end{equation}
As in THVR\cite{ERTD2HI}, by two-hop knowledge, node $x$ can calculate the velocity offered by each of the two-hop favorable forwarding pairs ($F_1^{+p}(x)$,$F_2^{+p}(x)$) as shown in Eqn. \ref{eqn:Eq9}. Furthermore, we include queuing delay at both the current ($dq_x)$) and the next hop ($dq_y$) nodes, with the two-hop transmission delay ($dt_{xy}$ and $dt_{yz}$), this distinguishes the proposed protocol
from LOCALMOR[5].
\begin{equation}
\begin{tiny}
\begin{split}
V_{xy\rightarrow z} = \frac{dist(x,D) - dist(z,D)}{dq_x<pak.class> + dt_{xy} + dq_y<pak.class> + dt_{yz}}
\end{split}
\label{eqn:Eq9}
\end{tiny}
\end{equation}
Where, $y \in F_1^{+p}(x)$ and $z \in F_2^{+p}(x)$. The required velocity is relative to the progress made towards the destination
[21] and the time remaining to the deadline, $lt$ (lag time). The lag time is the time remaining until the packet deadline expires. At each hop, the transmitter renews this parameter in the packet header $i.e.$,
\begin{equation}
lt = lt_{p} - (t_{tx} - t_{rx} + sizeof(packet)/BW)
\label{eqn:Eq10}
\end{equation}
Where $lt$ is the time remaining to the deadline ($t_{req}$), $lt_{p}$ is the previous value 
of $lt$, ($t_{tx} - t_{rx} + sizeof(packet)/BW$) accounts for the delay 
from reception of the packet until transmission. On reception of the packet the node $x$, uses $lt$ to calculate the 
required velocity $V_{req}$ for all nodes in ($F_1^{+p}(x)$,$F_2^{+p}(x)$) as shown in Eqn. \ref{eqn:Eq11}.
\begin{equation}
V_{req} = \frac{dist(x,D)}{lt}
\label{eqn:Eq11}
\end{equation}
The node pairs satisfying $V_{xy\rightarrow z} \geq V_{req}$ form the set of nodes $S_{req}$, if the packet class
is $delay.responsive$ then the node with the maximum residual energy and minimum transmission power cost is chosen from the
set $S_{req}$. However, if the packet class is critical then the node with the highest packet reception ratio (PRR) is
selected from $S_{req}$, but if more than one node has the same maximum PRR, a node with maximum power efficiency is picked.

The Traffic-Differentiated Two-Hop Routing is shown in Algorithm 1, the computation complexity of this 
algorithm is $O(F_2^{+p}(x))$. Our proposed protocol is different from LOCALMOR, as it considers
two-hop neighborhood information that will provide enhanced foresight to the sender in identifying the node that can 
offer the required QoS and route the packets in real-time.

\begin{algorithm}
\begin{footnotesize}
\KwIn{$Packet$, $Queues$}
\BlankLine
\For{each Packet in node}{
	\If{packet.class == critical}{
		Place Packet in Critical-Queue;
	}
	\Else{
		\If{packet.class == delay.responsive}{
			Place Packet in Delay-Responsive-Queue;
		}
		\Else{
			Place Packet in Reliability-Responsive-Queue;
		}
		Start Timer;
	}
}
\For{each Timer Expire}{
	Shift packet to Critical Queue;
}
\For{each Packet Transmission}{
	Stop Timer;
}
\caption{Queuing Controller}
\label{algo:qm}
\end{footnotesize}
\end{algorithm}

\subsection{Queuing Controller}
The queuing controller helps accomplish low delay when routing critical and delay-responsive packets, higher precedence 
should be given to these packets in channel contention than the normal packets (regular and reliability-responsive packets). 
Additionally, critical packets need higher priority than delay-responsive
packets. This can be accomplished by implemented the queuing controller
module as detailed in Algorithm 2 [5]. Three queues are used to send packets from the highest priority queue to the lowest one. 
The highest priority queue, Critical-Queue, is used by critical packets, the second highest priority queue, Delay-Responsive-Queue,
is used by delay-responsive packets, and the least priority queue,
Reliability-Responsive-Queue, is used by regular and reliability-responsive packets. The
number of critical and delay-responsive packets is usually small, and there would be instances where their corresponding queues
are vacant, else higher priority traffic may hinder lower priority traffic.
Hence, a timer for each packet is employed to move it to the highest priority queue.

\section{Performance Evaluation}

To evaluate the proposed protocol, we carried out a simulation study using $ns$-$2$ [22]. In this study the proposed protocol 
(LRDTHR) is compared with LOCALMOR, DARA and MMSPEED. The simulation configuration consists of 900 nodes 
located in a 1800 $m^2$ area. Nodes are distributed following Poisson point process with a node density of 0.00027 node/$m^2$. The 
primary and secondary sink nodes are located in the region (0,0) and (1800, 1800) while the source node is located in the center
of the simulation area, equidistant from both the sinks. The source generated a CBR flow of 1 kB/second with a packet size of 
150 bytes. Critical and regular packets are used in the simulation for comparing our protocol with LOCALMOR, DARA and MMSPEED, 
while delay-sensitive and reliability-sensitive packets are used for comparing with LOCALMOR only. The deadline requirement was 
fixed in this simulation to 300ms for all class of packets.
\vskip 2mm
The MAC layer, link quality and energy consumption parameters are
set as per TelosB$^{3}$ (TPR2420) mote [23] with CC2420 radio as per LOCALMOR. Table 3 summarizes the simulation parameters.
LOCALMOR, DARA and MMSPEED are QoS protocols and a comparison of PRR (Packet Reception Ratio), 
ECPP (Energy Consumed Per Packet $i.e.$, the total energy expended divided by the number of packets effectively transmitted), 
packet average end-to-end delay (mean of packet delay) and the network lifetime are obtained.

\begin{table}[ht]
\centering
\caption{Simulation Parameters.}
\begin{tiny}
\begin{tabular}{|l|c| } \hline
{\bf Simulation Parameters} & {\bf  Value} \\ \hline
{Number of nodes} 		&{900} \\ \hline
{Simulation Topology} 		&{1800m x 1800m} \\ \hline
{Traffic} 			&{CBR} \\ \hline
{Critical Packet Rate} 		&{From 0 to 1} \\ \hline
{Regular Packet Rate} 		&{1 - CPR} \\ \hline
{Payload Size} 			&{150 Bytes} \\ \hline
{Transmission Power Range}			&{100m} \\ \hline
{Initial Battery Energy }		&{2.0 Joules} \\ \hline
{Energy Consumed during Transmit }	&{0.0522 Joule} \\ \hline
{Energy Consumed during Receive }	&{0.0591 Joule} \\ \hline
{Energy Consumed during Sleep }		&{0.00006 Joule} \\ \hline
{Energy Consumed during Idle }		&{0.000003 Joule} \\ \hline
{Propagation Model}		&{Free Space} \\ \hline
{Hello Period} 			&{5 seconds} \\ \hline
{PRR - WMEWMA Window} 			&{30} \\ \hline
{PRR - WMEWMA Weight Factor ($\beta$)}	&{0.6} \\ \hline
{Queuing/Delay - EWMA Weight Factor ($\gamma$)}	&{0.5} \\ \hline
\end{tabular}
\end{tiny}
\end{table}

In the first set of simulations the critical packet rate was varied from 0.1 to 1 and the remaining rate to 1 represents 
regular packet rate. Figure 1 and Figure 2 illustrates the efficiency of the TDTHR algorithm in increasing the PRR with respect
to regular and critical packets.
TDTHR, LOCALMOR and DARA linearly increase their performance as a function
of critical packet rate, while performance of MMSPEED is relatively constant. 
The high reliability of TDTHR, LOCALMOR and DARA is due to the use of efficient re-duplication towards different sinks, compared to MMSPEED that uses a multi-path single-sink approach, this results in packet congestion either at the final
sink or intermediate nodes.

\begin{figure}
\centering
\includegraphics[width=16pc, height=11pc]{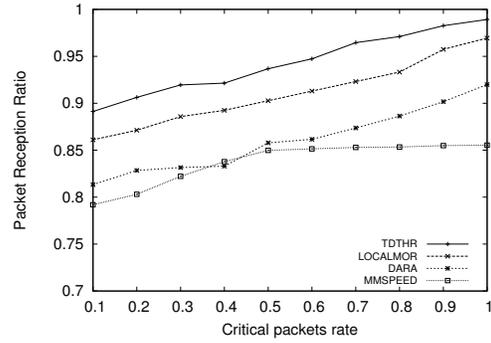}
\caption{Packet Reception Ratio - Regular Packets}
\end{figure}

\begin{figure}
\centering
\includegraphics[width=16pc, height=11pc]{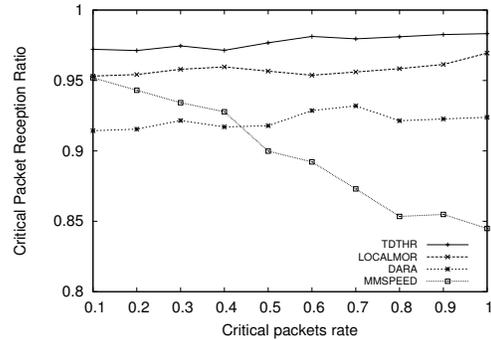}
\caption{Packet Reception Ratio - Critical Packets}
\end{figure}
\vskip 2mm
In Figure 2 the linear increase of the
packet reception ratio for TDTHR, LOCALMOR and DARA with the
increasing critical packet rate can be explained by the
subsequent increase of re-duplications (addressed only to critical
packets). Hence, the larger the number of critical
packets we have, the more the packets are duplicated,
which eventually increases their reception ratio.
In TDTHR the two-hop based routing and dynamic velocity of the TDTHR algorithm is able to aggressively route more packets
to the sink node, hence it is observed that TDTHR has higher PRR than the others in general.

\begin{figure}
\centering
\includegraphics[width=16pc, height=11pc]{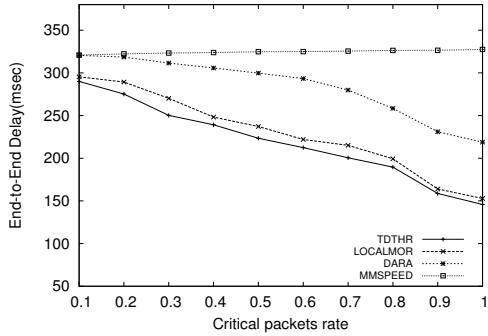}
\caption{End-to-End Delay - Regular Packets}
\end{figure}

\begin{figure}
\centering
\includegraphics[width=16pc, height=11pc]{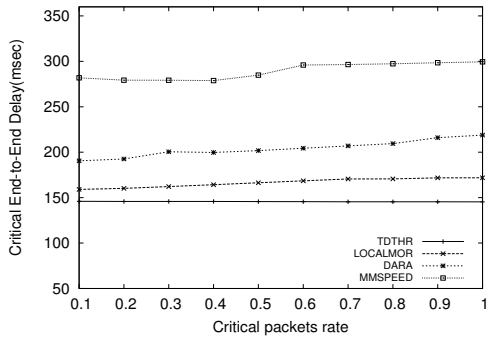}
\caption{End-to-End Delay - Critical Packets}
\end{figure}
\vskip 2mm
Figure 3 and Figure 4 illustrate the packet end-to-end delay of regular packets and critical packets respectively, performance 
of LRTHR is better that the other protocols.
LOCALMOR and DARA consider one-hop transmission delay and queuing waiting time, while TDTHR considers
dynamic velocity, two-hop transmission delay and queuing waiting time hence the selected paths from 
source to sink will be shorter and aid in reducing the end-to-end delay.
MMSPEED also considers queuing and transmission delays, but on the other hand, the use of multi-path single-sink 
transmissions causes congestion and thus results in several retransmission of packets before successful reception, which 
explains the relatively higher delay. In Figure 4 we notice a stable end-to-end delay for all protocols, which indicates
the strength of the routes selected for critical packets that are clearly not affected by the rise in rate of critical packets.

\begin{figure}
\centering
\includegraphics[width=16pc, height=11pc]{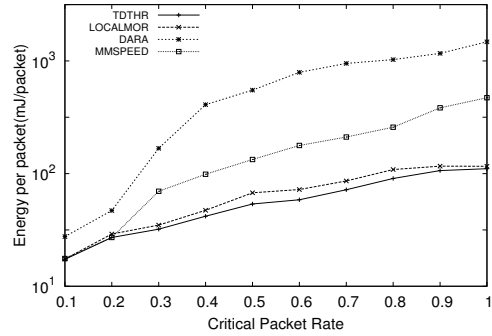}
\caption{ECPP vs Packet Rate}
\end{figure}

\begin{figure}
\centering
\includegraphics[width=16pc, height=11pc]{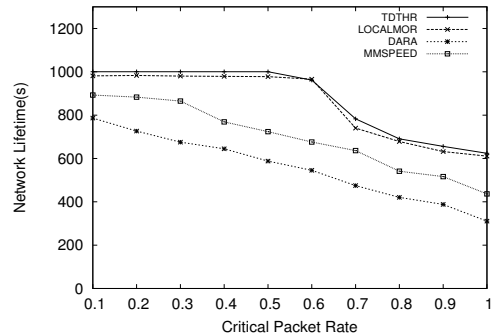}
\caption{Network Lifetime vs Packet Rate}
\end{figure}
\vskip 2mm
As depicted in Figure 5 the energy consumption per packet (ECPP) successfully transmitted, ascend as the 
critical packet rate increases.
The energy consumption has similar tendency in both TDTHR and LOCALMOR but DARA has a higher energy utilization. 
In LOCALMOR the energy consumption per packet smoothly increases as packet rates become higher.
LOCALMOR balances the load among nodes that ensure delivery within the deadline and have high reliability.
DARA performs inadequately in terms of energy, as it does neither use any traffic balancing approach nor any probabilistic
allocation. 
\vskip 2mm
In TDTHR the two-hop based routing will ensure shorter paths between source and sink, by selecting links providing 
higher PRR on the route to the sink, the energy consumption of the forwarding nodes can be 
minimized, due to lower number of collisions and re-transmissions and help in traffic balancing. 
Furthermore, in the proposed protocol the link delay and 
packet delivery ratios are updated by piggybacking the information in ACK, this will help in reducing the number of feedback
packets and hence reduce the total energy consumed. The impact of efficient energy utilization and traffic balancing on network lifetime is depicted in Figure 6, TDTHR and LOCALMOR show good performance compared to DARA and MMSPEED.
\vskip 2mm
Last, we study the performance of TDTHR and LOCALMOR with respect to delay-responsive and reliability-responsive traffic.
The QoS traffic is varied in the same way as critical packets were varied in the earlier simulations, $i.e.,$ each QoS 
traffic varies from 0.1 to 1. Figure 7 and Figure 8 examines the results.
This comparison is important because we need to ascertain the positive effect of two-hop delay incorporated in TDTHR over
the one-hop delay used in the LOCALMOR. The delay-responsive traffic are routed through more delay efficient
links, while reliability-responsive traffic, considers only reliable links. From Figure 7 and Figure 8 it is clear that
the performance of TDTHR is better than LOCALMOR for delay-responsive traffic due to
two-hop information and has similar performance for reliability-responsive traffic.

\begin{figure}
\centering
\includegraphics[width=16pc, height=11pc]{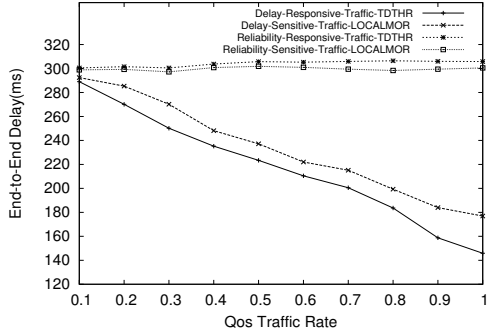}
\caption{TDTHR vs LOCALMOR - End-to-End delay}
\end{figure}

\begin{figure}
\centering
\includegraphics[width=16pc, height=11pc]{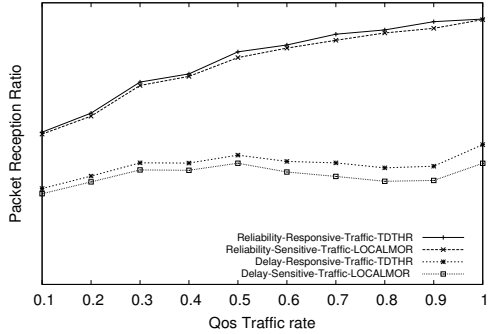}
\caption{TDTHR vs LOCALMOR - Packet Reception Ratio}
\end{figure}

\section{Conclusions}
\label{section:conclusions}

In this paper, we propose a Traffic-Differentiated Two-Hop Routing protocol for quality of service (QoS) in WSNs, it provides a
differentiation routing using different quality of service metrics.
Data traffic has been sequenced into different classes according to the required QoS, where different routing metrics and techniques are used for each class. 
The protocol is able to augment real-time delivery by an able integration of multi-queue priority policy, 
link reliability, two-hop information and dynamic velocity. The protocol is able to increase the PRR, end-to-end delay and improve
the energy efficiency throughout the network. This makes the protocol suitable for WSNs with varied traffic, such
as medical and vehicular applications.

\small
\balance

\noindent{\includegraphics[width=1in,height=1.7in,clip,keepaspectratio]{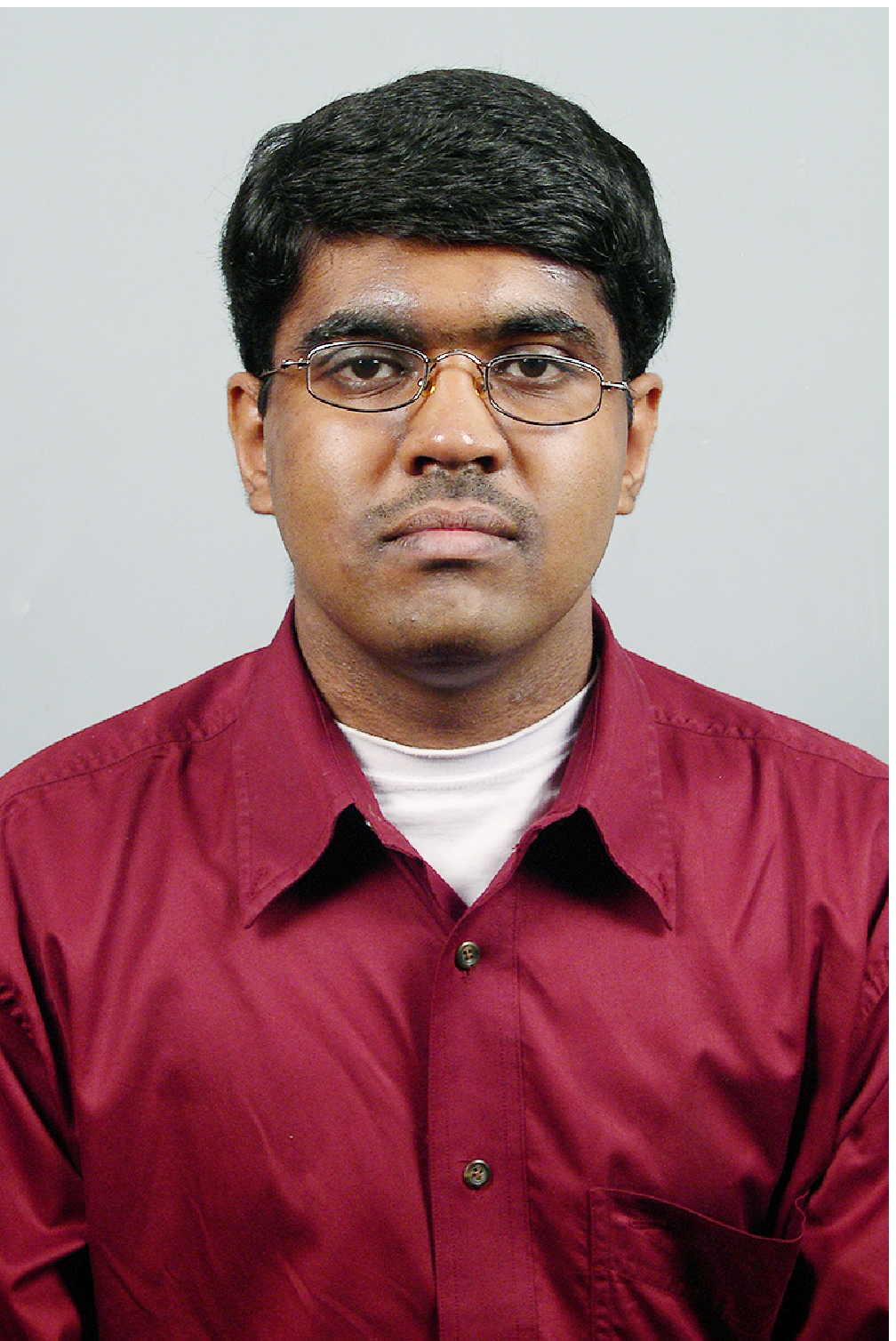}}
\begin{minipage}[b][1in][c]{1.8in}
{\centering{\bf {T Shiv Prakash}} is an Assistant Professor in the Department of Computer Science and Engineering at Vijaya Vittala Institute of Technology, Bangalore, India. He obtained his B.E and M.S Degrees in Computer Science and Engineering from Bangalore University, Bangalore. He is presently pursuing his Ph.D. pr-}\\\\\\
\end{minipage}\\
ogramme in the area of Wireless Sensor Networks in Bangalore University. His research interest is in the area of Sensor Networks, Embedded Systems and Digital Multimedia. \\\\

\noindent{\includegraphics[width=1in,height=1.7in,clip,keepaspectratio]{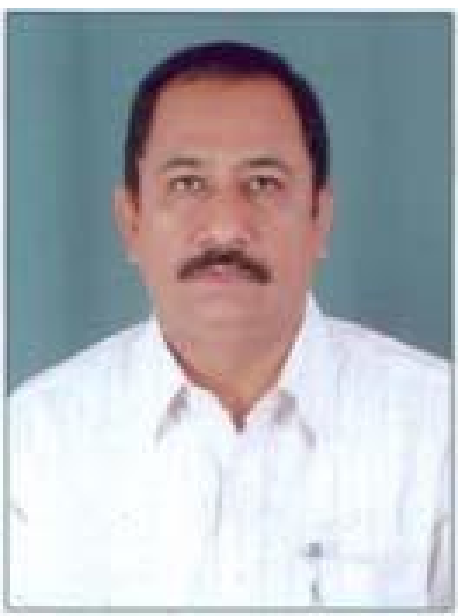}}
\begin{minipage}[b][1in][c]{1.8in}
{\centering{\bf {K B Raja }} is an Associate Professor, Dept. of Electronics and Communication Engg, University Visvesvaraya college of Engg, Bangalore University, Bangalore. He     obtained his Bachelor of Engineering and Master of Engineering in Electronics and Communication  En- }\\\\
\end{minipage}
gineering from University   Visvesvaraya  College     of Engineering, Bangalore. He was awarded
Ph.D. in Computer Science and Engineering from Bangalore
University. He has over 100 research publications in refereed
International Journals and Conference Proceedings. His research
interests include Image Processing, Biometrics, VLSI      Signal
Processing, Computer Networks.
 \\\\
\noindent{\includegraphics[width=1in,height=1.8in,clip,keepaspectratio]{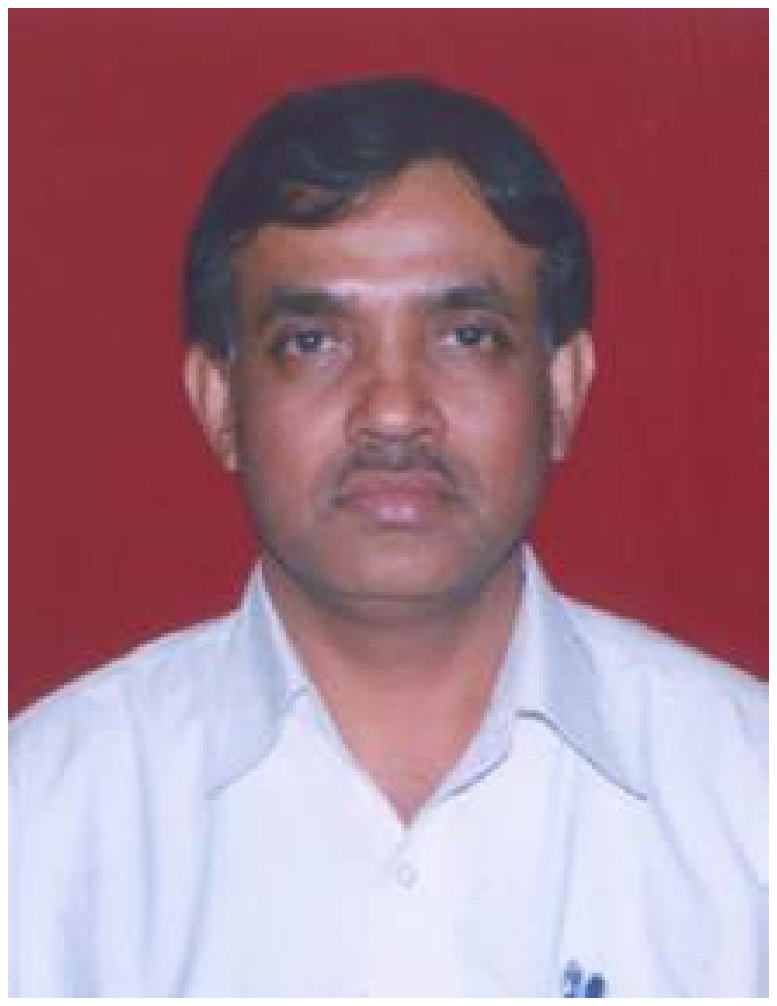}}
\begin{minipage}[b][1in][c]{1.8in}
{\centering{\bf {Venugopal K R}} is currently the Principal, University Visvesvaraya College of Engineering, Bangalore University, Bangalore. He obtained his Bachelor of Engineering from University Visvesvaraya College of Engineering. He received his Masters degree in Computer Science and} \\ \\
\end{minipage}
Automation from Indian Institute of Science Bangalore. He was awarded Ph.D. in Economics from Bangalore University and Ph.D. in Computer Science from Indian Institute of Technology, Madras. He has a distinguished academic career and has degrees in Electronics, Economics, Law, Business Finance, Public Relations, Communications, Industrial Relations, Computer Science and Journalism. He has authored and edited 39 books on Computer Science and Economics, which include Petrodollar and the World Economy, C Aptitude, Mastering C, Microprocessor Programming, Mastering C++ and Digital Circuits and Systems etc.. During his three decades of service at UVCE he has over 350 research papers to his credit. His research interests include Computer Networks, Wireless Sensor Networks, Parallel and Distributed Systems, Digital Signal Processing and Data Mining.\\ \\

\noindent{\includegraphics[width=1in,height=1.6in,clip,keepaspectratio]{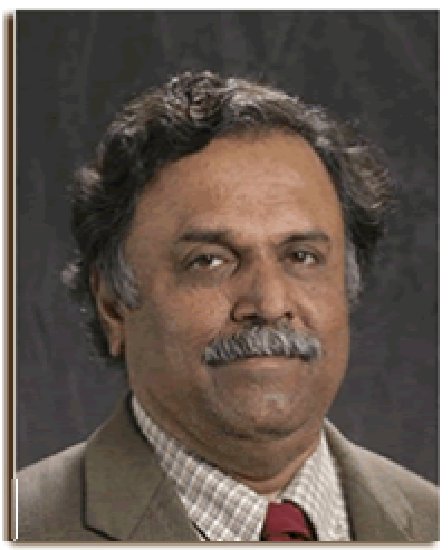}}
\begin{minipage}[b][1in][c]{1.8in}
{\centering{\bf {S S Iyengar}} is currently the Roy Paul Daniels Professor and Chairman of the Computer Science Department at Louisiana State University. He heads the Wireless Sensor Networks Laboratory and the Robotics Research Laboratory at LSU.}\\\\
\end{minipage}
He has been involved with research in High Performance Algorithms, Data Structures, Sensor Fusion and Intelligent Systems, since receiving his Ph.D degree in 1974 from MSU, USA. He is Fellow of IEEE and ACM. He has directed over 40 Ph.D students and 100 Post Graduate students, many of whom are faculty at Major Universities worldwide or Scientists or Engineers at National Labs/Industries around the world. He has published more than 500 research papers and has authored/co-authored 6 books and edited 7 books. His books are published by John Wiley \& Sons, CRC Press, Prentice Hall, Springer Verlag, IEEE Computer Society Press etc.. One of his books titled Introduction to Parallel Algorithms has been translated to Chinese.\\\\

\noindent{\includegraphics[width=1in,height=1.8in,clip,keepaspectratio]{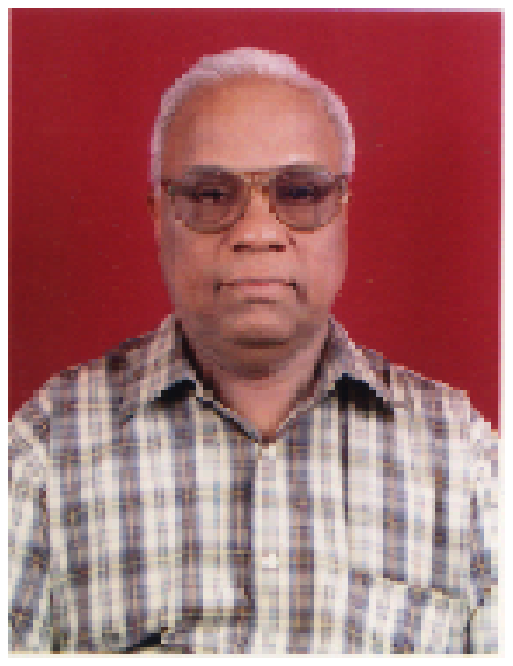}}
\begin{minipage}[b][1in][c]{1.8in}
{\centering{\bf {L M Patnaik}} is currently Honorary Professor, Indian Institute of Science, Bangalore, India. He was a Vice Chancellor, Defense Institute of Advanced Technology, Pune, India and was a Professor since 1986 with the Department of Computer Science and Automation, Indian} \\ \\
\end{minipage}
Institute of Science, Bangalore. During the past 35 years of his service at the Institute he has over 700 research publications in refereed International Journals and refereed International Conference Proceedings. He is a Fellow of all the four leading Science and Engineering Academies in India;  Fellow of the IEEE and the Academy of Science for the Developing World. He has received twenty national and international awards; notable among them is the IEEE Technical Achievement Award for his significant  contributions to High Performance Computing and Soft Computing. His areas of research interest have been Parallel and Distributed Computing, Mobile Computing, CAD for VLSI circuits, Soft Computing and Computational Neuroscience.
\balance

\begin{thebibliography}{37}





\bibitem{LCDJ}
{F  L Lewis, D J Cook, S K Dasm and John Wiley}. {Wireless Sensor
  Networks}, in \emph{Proc. Smart Environment Technologies, Protocols and
  Applications, New York},  pages 1--18. 2004.

\bibitem{MMSPEED1}
{E Felemban, C G Lee and E Ekici}. {MMSPEED: Multipath Multi-Speed
  Protocol for QoS Quarantee of Reliability and Timeliness in Wireless Sensor
  Network}, \emph{in IEEE Transactions on Mobile Computing}, 5(6): 738--754, 2006.

\bibitem{DARA1}
{M M Or-Rashid, Md. Abdur Razzaque, M M Alam and C S Hong}.
  {Multi-Constrained QoS Geographic Routing for Heterogeneous Traffic in
  Sensor Networks}, \emph{in IEICE Transactions on Communications}, 91B(8):2589--2601, 2008.

\bibitem{ERTD2HI}
{Y Li, C S Chen and Y Q Song}. {Enhancing Real-Time Delivery In Wireless
  Sensor Networks With Two-Hop Information}, \emph{in IEEE Transactions On
  Industrial Informatics}, 5(2):113--122, 2009.

\bibitem{LOCALMOR1}
{Djamel Djenouri and Ilangko Balasingham}. {Traffic-Differentiation-Based
  Modular QoS Localized Routing for Wireless Sensor Networks}, \emph{in IEEE
  Transactions on Mobile Computing}, 6(10):797--809, 2011.

\bibitem{SPEED1}
{Tian He, John A Stankovic, Chenyang Lu and Tarek F Abdelzaher}. {A
  Spatiotemporal Protocol for Wireless Sensor Network}, \emph{in IEEE
  Transactions on Parallel and Distributed Systems}, 16(10):995--1006, 2005.

\bibitem{GPSR1}
{B Karp and Kung H T}. {GPSR: Greedy Perimeter Stateless Routing for
  Wireless Networks}, in \emph{Proc. 6th Annual International Conference on
  Mobile Computing and Networking (MobiCom)},  pages 243--254, 2000.

\bibitem{RGDADHOC}
{Prosenjit Bose, Pat Morin, Ivan Stojmenovi and Jorge Urrutia}, {Routing
  with Guaranteed Delivery in Ad hoc Wireless Networks}, in \emph{Proc. of
  3rd ACM Int. Workshop on Discrete Algorithms and Methods for Mobile Computing
  and Communications DIALM'99},  pages 48--55, Aug. 1999.

\bibitem{PIAQC}
{A Sharif, V Potdar and A J D Rathnayaka}. {Prioritizing Information for
  Achieving QoS Control in WSN}, in \emph{Proc. IEEE International Conference
  on Advanced Information Networking and Applications},  pages 835--842, 2010.

\bibitem{MCAMOR}
{M E Rusli, R Harris and A Punchihewa}. {Markov Chain-based analytical
  model of Opportunistic Routing protocol for wireless sensor networks}, in
  \emph{Proc. TENCON IEEE Region 10 Conference},  pages 257--262, 2010.

\bibitem{QDGRP1}
{M Koulali, A Kobbane, M El Koutbi and M Azizi}. {QDGRP: A Hybrid QoS
  Distributed Genetic Routing Protocol for Wireless Sensor Networks}, in
  \emph{Proc. International Conference on Multimedia Computing and Systems},
   pages 47--52, 2012.

\bibitem{CLAE2EDD}
{Yunbo Wang, M C Vuran and S Goddard}. {Cross-Layer Analysis of the
  End-to-End Delay Distribution in Wireless Sensor Networks}, \emph{in IEEE
  Transactions on Networking}, 20(1):305--318, 2012.

\bibitem{RMACAR}
{S Ehsan, B Hamdaoui and M Guizani}, {Radio and Medium Access Contention
  Aware Routing for Lifetime Maximization in Multichannel Sensor Networks},
  \emph{in IEEE Transactions on Wireless Communication}, 11(9):3058--3067, 2012.

\bibitem{GREES1}
{K Zeng, K Ren, W Lou and P J Moran}. {Energy Aware Efficient
  Geographic Routing in Lossy Wireless Sensor Networks with Environmental
  Energy Supply}, \emph{in Wireless Networks}, 15(1):39--51, 2009.

\bibitem{DHGR1}
{M Chen, V Leung, S Mao, Y Xiao, and I Chlamtac}. {Hybrid Geographical
  Routing for Flexible Energy-Delay Trade-Offs}, \emph{in IEEE Transactions
  on Vehicular Technology}, 58(9):4976--4988, 2009.

\bibitem{EAGFS1}
{T L Lim and M Gurusamy}. {Energy Aware Geographical Routing and Topology
  Control to Improve Network Lifetime in Wireless Sensor Networks}, in
  \emph{Proc. IEEE International Conference on Broadband Networks
  (BROADNETS’05)},  pages 829--831, 2005.

\bibitem{EGRKH}
{C S Chen, Y Li and Y Q Song}. {An Exploration of Geographic Routing
  with K-hop Based Searching in Wireless Sensor Networks}, in \emph{Proc.
  CHINACOM},  pages 376--381, 2008.

\bibitem{LRTHR1}
{T Shiva Prakash, K B Raja, K R Venugopal, S S Iyengar, L M
  Patnaik}. {Link-Reliability Based Two-Hop Routing for QoS Guarantee in
  Wireless Sensor Networks}, in \emph{IEEE Proc. of the 16th International
  Symposium on Wireless Personal Multimedia Communications}, 2013.

\bibitem{RFLLSSN}
{T He, C Huang, B M Blum, J A Stankovic and T F Abdelzaher},
  {Range-Free Localization and Its Impact on Large Scale Sensor Networks},
  \emph{in ACM Trans. Embedded Computer Systems}, 4(4): 877--906, 2000.

\bibitem{EELRE1}
{A Woo and Culler}. {Evaluation of Efficient Link Reliability Estimators for
  Low-Power Wireless Networks}, University of California, Tech. Rep., 2003.

\bibitem{RPAR1}
{O Chipara, Z He, G Xing, Q Chen, X Wang, C Lu, J Stankovic and T
  Abdelzaher}, {Real-Time Power-Aware Routing in Sensor Network}, in
  \emph{Proc. IWQoS}, , pages 83--92, Jun. 2006.

\bibitem{36}
NS-2, [Online]. Available: {http://www.isi.edu/ nsnam/ns/}.

\bibitem{37}
Crossbow Motes, [Online]. Available: {http:// www.xbow.com}.\\\\
\small
\balance
\end{thebibliography}
\end{document}